\documentclass[journal=nalefd,manuscript=letter]{achemso}
\usepackage[utf8]{inputenc}
\usepackage[T1]{fontenc}
\usepackage[english]{babel}
\usepackage{graphicx}
\usepackage{amsmath}
\usepackage{color}
\usepackage{ulem}
\usepackage{gensymb}
\usepackage{latexsym}
\usepackage{amssymb}
\usepackage{multirow}
\usepackage{booktabs}
\usepackage[version=4]{mhchem}
\usepackage[]{siunitx}
\usepackage{amssymb}

\newcommand{\csix}{\ce{C60}}
\newcommand{\bite}{\ce{Bi4Te3}}

\newcommand*{\TUW}{Institut f\"ur Festk\"orperelektronik, Technische Universit\"at Wien, Gu{\ss}hausstra{\ss}e 25, 1040 Vienna, Austria }

\usepackage{amsmath}
\let\oldAA\AA
\renewcommand{\AA}{\text{\normalfont\oldAA}}

\title{Molecular order induced charge transfer in a C$_{60}$-topological insulator moir\'e heterostructure}

\author{Ram Prakash Pandeya}
\email{ram.pandeya@tuwien.ac.at}
\affiliation{\TUW}
\altaffiliation{II. Physikalisches Institut, Universit\"at zu K\"oln, Z\"ulpicher Strasse 77, 50937 K\"oln, Germany}

\author{Konstantin P. Shchukin}
\affiliation{\TUW}
\altaffiliation{II. Physikalisches Institut, Universit\"at zu K\"oln, Z\"ulpicher Strasse 77, 50937 K\"oln, Germany}

\author{Yannic Falke}
\affiliation{II. Physikalisches Institut, Universit\"at zu K\"oln, Z\"ulpicher Strasse 77, 50937 K\"oln, Germany}

\author{Gregor Mussler}
\affiliation{Peter Gr\"unberg Institut (PGI-9), Forschungszentrum J\"ulich, D-52425 J\"ulich, Germany}

\author{Jalil Abdur Rehman}
\affiliation{Peter Gr\"unberg Institut (PGI-9), Forschungszentrum J\"ulich, D-52425 J\"ulich, Germany}

\author{Nicolae Atodiresei}
\affiliation{Peter Gr\"unberg Institut (PGI-1), Forschungszentrum J\"ulich, D-52425 J\"ulich, Germany}

\author{Alexander V. Fedorov}
\affiliation{II. Physikalisches Institut, Universit\"at zu K\"oln, Z\"ulpicher Strasse 77, 50937 K\"oln, Germany}

\author{Boris V. Senkovskiy}
\affiliation{II. Physikalisches Institut, Universit\"at zu K\"oln, Z\"ulpicher Strasse 77, 50937 K\"oln, Germany}

\author{Daniel Jansen}
\affiliation{II. Physikalisches Institut, Universit\"at zu K\"oln, Z\"ulpicher Strasse 77, 50937 K\"oln, Germany}

\author{Giovanni Di Santo}
\affiliation[Elettra]{Elettra Sincrotrone Trieste, Strada Statale 14 km 163.5, 34149 Trieste, Italy}

\author{Luca Petaccia}
\affiliation[Elettra]{Elettra Sincrotrone Trieste, Strada Statale 14 km 163.5, 34149 Trieste, Italy}

\author{Alexander Gr\"uneis}
\email{alexander.grueneis@tuwien.ac.at}
\affiliation{\TUW}
\altaffiliation{II. Physikalisches Institut, Universit\"at zu K\"oln, Z\"ulpicher Strasse 77, 50937 K\"oln, Germany}

\begin{document}
\begin{abstract}
  We synthesize and spectroscopically investigate monolayer \csix{} on the topological insulator (TI) \bite{}. This \csix{}/\bite{} heterostructure is characterized by excellent translational order in a novel ($4\times4$) \csix{} superstructure on a ($9\times9$) unit of \bite{}. We measure the full two-dimensional energy band structure of \csix{}/\bite{} using angle-resolved photoemission spectroscopy (ARPES). We find that \csix{} accepts electrons from the TI at room temperature but no charge transfer occurs at low temperatures. We unravel this peculiar behaviour by Raman spectroscopy of \csix{}/\bite{} and density functional theory (DFT) calculations of the electronegativity of \csix{}. Both methods are sensitive to orientational order of \csix{}. At low temperatures, Raman spectroscopy shows a dramatic intensity increase of the \csix{} Raman signal, evidencing a transition to a rotationally ordered state. DFT reveals that the orientational order of \csix{} at low temperatures has a higher electron affinity than at high temperatures. These results neatly explain the temperature-dependent charge transfer observed in ARPES. Our conclusions are supported by the appearance of a strong photoluminescence from \csix{}/\bite{} at low temperatures. \\
\end{abstract}

\maketitle

\section{Introduction}
Organic thin films deposited on metal or semiconductor substrates can form interfaces with exciting opto-electronic properties that are controlled by the charge transfer between the molecule and the substrate and the molecular order. The large variety of organic molecules allows for a flexible modification of the interface properties.~\cite{Chugh2019} Two intertwined aspects for obtaining interfaces with defined properties are the molecular order and the molecule-substrate-interaction.~\cite{Gruznev2013Apr} A strong molecule-substrate interaction can imprint molecular order, however it is detrimental to inheriting the molecular properties to the interface electronic states because it alters the molecular character of the electronic states at the surface. Thus, one wishes to fine-tune the strength of the molecule-substrate interaction versus molecule-molecule interaction by choice of the substrate and molecule. The surfaces of bulk metals and also semiconductors often form covalent bonds to molecules because dangling bonds are present on their surface. Moreover, the interface structure is complicated by surface reonstructions, change in molecular shape and geometrical frustration.~\cite{AlfonsoMoro2023} E.g. electron doping of \csix{} on the reconstructed Cu surface is optimal for obtaining superconductivity, i.e. C$_{60}^{3+}$~(Ref.~\citenum{Pai2010}) and a large $7\times 7$ superstructure forms for \csix{} on Au.~\cite{Schull2007Nov} On the other hand, van-der-Waals substrates have weaker interaction to molecules due to absence of dangling bonds. For example, for pentacene on graphene, a coverage dependent molecular orientation has been oberserved that could further be tuned by doping of the graphene substrate.~\cite{Liu2013} Similarly, it has been observed that the charge transfer between \csix{} and graphene can be controlled by a gate voltage applied to graphene; this charge transfer in turn controls the growth mode and molecular order of \csix{}.~\cite{Nguyen2020} Unbiased graphene becomes p-doped when covered by \csix{} which has been revealed by an upshift in the Raman active 2D band of graphene and a shift of the conductance minimum as a function of gate voltage.~\cite{Qin2020,Matkovic2017} Van-der-Waals epitaxy and interaction of organic molecules has been oberserved for PTDCA on graphene~\cite{Wang2009}, \csix{} on hBN \cite{Muntwiler2005,Guo2020}, \csix{} on WSe$_2$ \cite{Santos2017}, PDCTA on borphene\cite{Liu2024}. Such heterostructures are excellent materials for vertical field effect transistors and opto-electronic devices.~\cite{Huang2020,Kim2015,Oswald2023,Kleemann2020,Choi2020,He2015}
Much less is known about the interaction of organic molecules with topological insulators (TIs). The organic layer/TI system allows for studying the stability of topologically protected electronic states and provides a new substrate for growth. To that end the molecule-TI interactions have been investigated for monolayers of MnPc, CoPc, and CuPc~\cite{Sessi2014Sep,Bathon2015Apr} and PTCDA.~\cite{Yang2015Oct} Highly ordered \csix{} in a hexagonal structure with a thickness of 50~\AA{} has been synthesized on Bi$_2$Se$_3$ topological insulator.~\cite{Latzke2019} Yet, the rather large thickness of the organic \csix{} layers prevented a study of the structure of the interface by surface science methods such as photoemission or scanning probe techniques. In particular, the charge transfer between the TI to the \csix{}, the change of the topologically protected surface state, the molecular orientation and the optical properties of the interface are not known. \csix{} stands out from other organic molecules as a highly symmetric molecule with a triply degnerate HOMO and LUMO level which exhibits strong electron-photon interaction and correlated electronic ground statee when alkali-metal doped.\cite{millie96} The \csix{} molecule acts as a cage for endohedral encapsulation of atoms along with a high degree of available functionalizations.\cite{Yan2014Dec, Li2015Feb, Carrillo-Bohorquez2021Sep, Komatsu2005Jan, Whitener2008Oct}  
For single crystal \csix{} and thick films of \csix{}, a structural phase transition occurs for \csix{} at 250~K from face centered cubic (fcc) to simple cubic (sc).\cite{millie96} This phase transition is related to the orientation of adjacent \csix{} molecules. For fcc, neighbouring molecules have different orientation whereas for sc there the orientation of neighouring \csix{} is identical.\cite{millie96}  This phase transition also has a large impact on the Raman spectra. That is, the intense Ag(2) mode, or pentagonal pinch mode hardens and becomes sharper below the transition temperature. It is clear that for a \csix{} monolayer on a substrate that phase transition can be modified or even absent. Yet, until now, there are no reports on the effect of substrate interaction on that transition despite it might be important for the charge transfer of \csix{} to the substrate.
Here, we investigate the properties of monolayer (ML) \csix{} on the novel topological insulator \bite{}.\cite{Yamana1979Jan, Chagas2020Mar, Chagas2022Feb, Nabok2022Mar} Interestingly, ML \csix{} grows in a ($4\times4$) \csix{} superstructure on a ($9\times9$) grid with regards to the underlying TI. This structure results in a sharp moir\'e pattern observed in low energy electron diffraction (LEED). Angle-resolved photoemission spectroscopy (ARPES) reveals that the TI becomes hole doped upon \csix{} deposition. Interestingly, the molecular order of \csix{} is crucial to the hole doping. That is, when the temperature is lowered to to 80~K by cooling with liquid N$_2$, the \csix{} molecules change the angle of orientation and the hole doping effect is reversed. Resonance Raman spectroscopy of the pentagonal pinch mode of \csix{} is shown to be highly sensitive to changes in molecular ordering: upon cooling to 80~K, that Raman mode increases dramatically in intensity and its linewidth narrows. Calculations using density functional theory (DFT) are able to explain the changes in charge transfer with the molecular orientation.

\section{Results}
\subsection{Electronic and structural properties of the \csix/\bite{} interface}
\bite{} condenses in the form of a rhombohedral crystal structure (R$\Bar{3}$m space group) with the
unit cell consisting of stacked quintuple layers (QL) with \ce{Bi$_2$Te$_3$} stoichiometry and Bi-Bi bilayers (BL).
The 20~nm Bi$_4$Te$_3$ thin films used in this study are grown in (0001) direction on Si(111) in ultra-high vacuum (UHV).~\cite{Nabok2022Mar} The (0001) surface of \bite{} has a hexagonal lattice with lattice constant $a_{\mathrm{hex}}=4.501$\,\AA{} as determined by X-Ray Diffraction (XRD) experiments.\cite{Yamana1979Jan} After growth, the surface of \bite{} is capped by Al$_2$O$_3$ which protects the surface against oxidation. Capped \bite{} samples can be brought to ambient conditions during transport from the molecular beam epitaxy (MBE) chamber to the angle-resolved photoemission spectroscopy (ARPES) system. After insertion into the ARPES system, the samples are decapped by sputtering and annealing (see methods).

\begin{figure*}[ht]
    \centering
    \includegraphics[width=17cm]{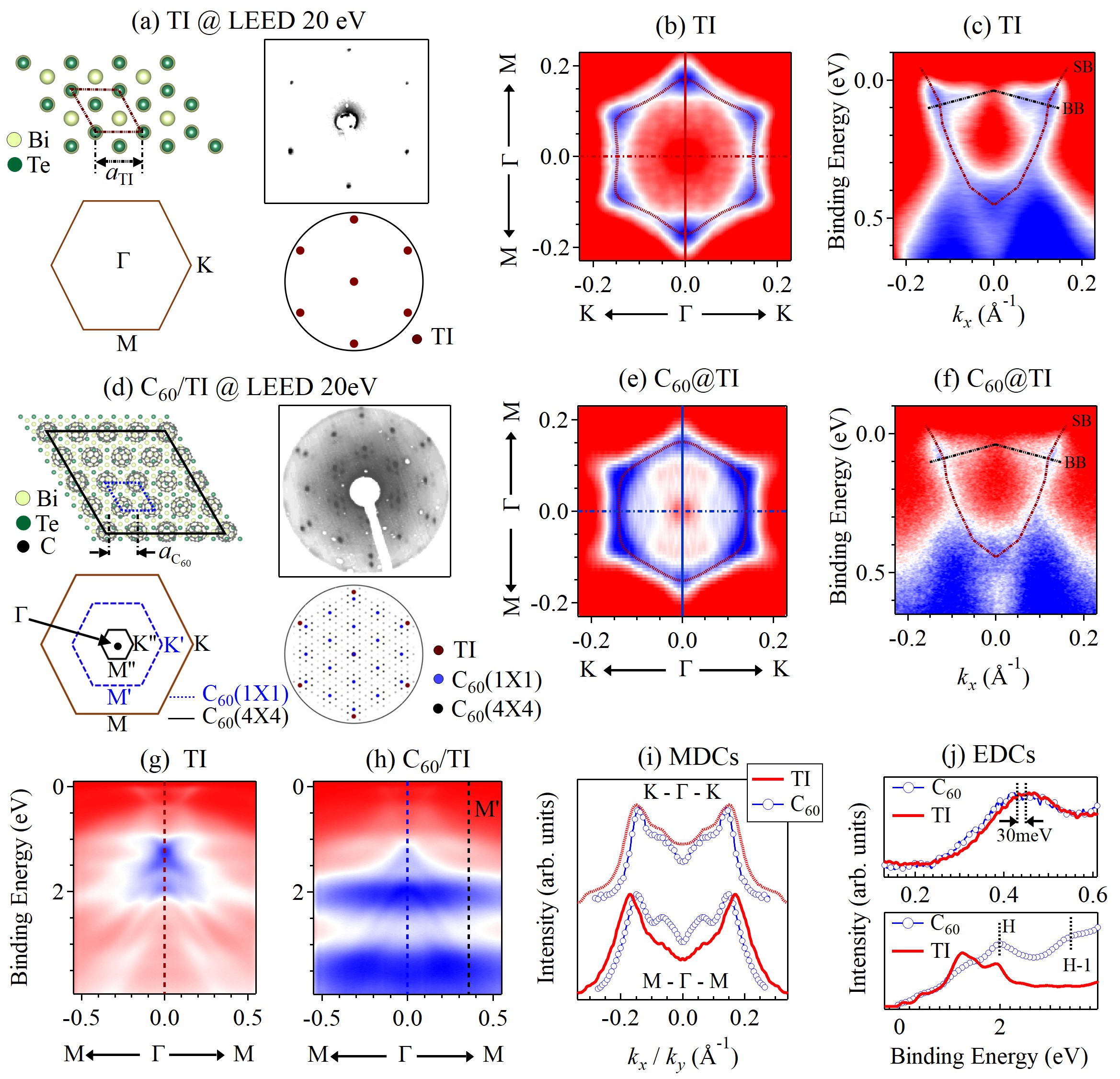}
    \caption{(a) Upper row: surface crystal structure and LEED pattern of \bite. Lower row: \bite{} BZ and sketch of the LEED pattern. (b) Fermi surface of \bite{} and tight-binding (TB) fit of the TI surface state (red). (c) ARPES of \bite{} along $\Gamma K$ direction. The surface bands (SB) and bulk bands (BB) are indicated. (d) Upper row: surface unit cell with the $4\times 4$ \csix{} superstructure on $9\times 9$ TI, the BZ and a sketch of the surface unit cell. Lower row: LEED pattern of monolayer \csix{} on \bite{}  and sketch of the LEED pattern reflexes from \bite{}, \csix{} and \csix{} superstructure. (e) Fermi surface of \csix{}/\bite{} along with a TB fit of the TI surface state (red). (f) ARPES scan of \csix{}/\bite{} along $\Gamma K$. (g) and (h) ARPES of \bite{} and \csix{}/\bite{} in a wide energy range including HOMO and HOMO-1 of \csix{} at 2~eV and 3.5~eV, respectively. (i) Momentum dispersion curves (MDCs) of \bite{} and \csix{}/\bite{} along $\Gamma K$ and $\Gamma M$. (j) Energy dispersion curves (EDCs) of \bite{} and \csix{}/\bite{} at $\Gamma$ in a narrow energy range around the Dirac point and in a wide energy range. The labels $H$ and $H-1$ indicate the HOMO and HOMO-1 levels. All ARPES data are taken at RT with $h\nu=21.2$~eV.}
    \label{fig:c60ti-leed}
\end{figure*}

The matching reciprocal space LEED of the pristine \bite{} thin film is shown in Fig.~\ref{fig:c60ti-leed}a along with a sketch of the observed LEED pattern, a sketch of the (0001) surface of \bite, and the first Brillouin zone. The sharp LEED reflexes in a hexagonal pattern indicate a clean surface and high crystallinity. This pattern is consistent over the whole sample area of $1\times1$\,cm$^2$, proving the absence of possible rotational domains of the surface. Figure~\ref{fig:c60ti-leed}b shows the hexagonally warped, n-type Fermi surface of \bite{} measured by ARPES along with a tight-binding (TB) fit to the Fermi surface.\cite{Fu2009} From the TB fit to ARPES, we determine a value of the ratio between hexagonal warping $\lambda$ to Fermi velocity $v$ of $\lambda/v=$ 27~\AA$^2$. The energy of the Dirac point is determined from ARPES maxima to be at 0.45~eV binding energy. Using the Fermi surface size, we estimated the carrier concentration to be $n = 8.98\times10^{12}$~cm$^{-2}$. Figure~\ref{fig:c60ti-leed}c depicts an ARPES scan through the BZ center along $\Gamma K$ direction in the vicinity of $E_F$. Though the bulk bands are visible close to the Fermi level, they appear not to cross the Fermi level.
  
Figure ~\ref{fig:c60ti-leed}d depicts the LEED pattern of one ML \csix{} evaporated on \bite{} along with a sketch of the LEED pattern and of the interface as well as the reciprocal lattice of the new supercell. The LEED pattern shows the original diffraction spots of \bite{} but in addition there are new diffraction spots with a slightly higher intensity that stem from \csix{}. From the LEED pattern we see that the reciprocal lattice vector that corresponds to the \csix{} is almost half of a \bite{} reciprocal lattice vector. Since the lattice constant of the \csix{} molecular layer is not close to an integer multiple of the lattice constant of \bite{}, a large moir\'e pattern forms. The set of diffraction spots of weak intensity around each \csix{} diffraction spot are due to that moir\'e pattern that \csix{} forms on the \bite{} surface. In the sketch of the LEED in the lower right panel of Figure~\ref{fig:c60ti-leed}d, the diffraction spots are colour-coded according to their origin. The pattern inherits the same orientation as the interface and is due to the arrangement of \csix{} molecules in a ($9\times9$) superstructure with respect to the TI lattice, as sketched in the upper left panel of Figure~\ref{fig:c60ti-leed}d. The new supercell contains ($4\times4=16$) \csix{} per unit cell and is therefore defined by a large  hexagonal lattice vector of about 4~nm length. The bottom left panel of Figure~\ref{fig:c60ti-leed}d depicts the BZ of \bite{} along with the \csix{} and the much smaller moir\'e BZs. Based on $a_{\bite{}}$, the lattice constant of \bite{}, we estimate the \csix{} lattice on \bite{} to be $(9/4)\times a_{\bite{}}$ = 10.15~\AA. The \csix{} lattice constant in the fcc structure is $a_{\csix}^{fcc}=14.12~\AA$.~\cite{sathish2012synthesis} Hence, the lattice constant of a hexagonal lattice defined on the fcc(111) plane is $a_{\csix{}}^{fcc}/\sqrt{2}$  which is equal to 10.01~\AA. This implies that the \csix{} monolayer on \bite{} is strained by $(10.15-10.01)/10.15\times 100=1.4$\%.

The electronic structure of the \csix{}/\bite{} interface is studied by ARPES and the resulting Fermi surface is shown in Figure~\ref{fig:c60ti-leed}e. Performing a TB fit with the same model as used in Figure~\ref{fig:c60ti-leed}b, a lower hexagonal warping of $\lambda/v= 22~\AA^2$ is obtained in addition to an upshifted Dirac point of $E=0.43$~eV. From the integrated Fermi surface area, we find that the carrier concentration is reduced to a value of $n$ = 7.59$\times10^{12}$ cm$^{-2}$, which is indicative of p-doping due to \csix{} deposition. Considering one ML \csix{} coverage, we estimate a charge transfer of 0.035 holes per \csix{} molecule. Figure~\ref{fig:c60ti-leed}f shows the ARPES scan of the \csix{}/\bite{} interface. The TI's surface state is still visible because we deposited just one ML of \csix{} through which photoelectrons coming from the TI can penetrate. In the wide energy range ARPES scans shown in Figures~\ref{fig:c60ti-leed}g and ~\ref{fig:c60ti-leed}h, the energy band structures of pristine \bite{} and \csix{}/\bite{} are compared. The obvious difference is the appearance of HOMO and HOMO-1 energy levels from \csix{} at binding energies of about 2~eV and 3.5~eV, respectively. In Figure~\ref{fig:c60ti-leed}i the momentum dispersion curves (MDCs) at the Fermi energy in $\Gamma K$ and $\Gamma M$ directions are shown for \bite{} and \csix{}/\bite{}. It can be clearly seen that the Fermi wavevector corresponding to the crossing of the surface state with the Fermi level is reduced for the \csix{}/\bite{} case in both the crystallographic directions. This is consistent with the observation of a reduction in Fermi surface size due to p-doping after \csix{} evaporation. Finally, in Figure~\ref{fig:c60ti-leed}j, we show the energy distribution curves (EDCs) of \bite{} and \csix{}/\bite{} at the $\Gamma$ point in a narrow energy region around the Dirac point and in a wide energy region that also covers HOMO and HOMO-1 levels. The direct comparison of EDCs close to Dirac point confirms shifting of the Dirac point towards the Fermi-level with \csix{} deposition. Our measurements also show the resilience of the TI surface state against perturbations induced by the \csix{} adsorbate layer. This resilience may be understood by the relatively small number of atomic sites on the TI surface that are in direct contact with the spherical \csix{} molecule. The \bite{} surface has Te atoms on its surface. The \csix{} molecules affect $4\times4=16$ out of $9\times9=81$ Te atoms in the superstructure unit cell. 
  
\subsection{\csix{} derived electron energy band structure of \csix/\bite}
\begin{figure*}[ht]
    \centering
    \includegraphics[width=17cm]{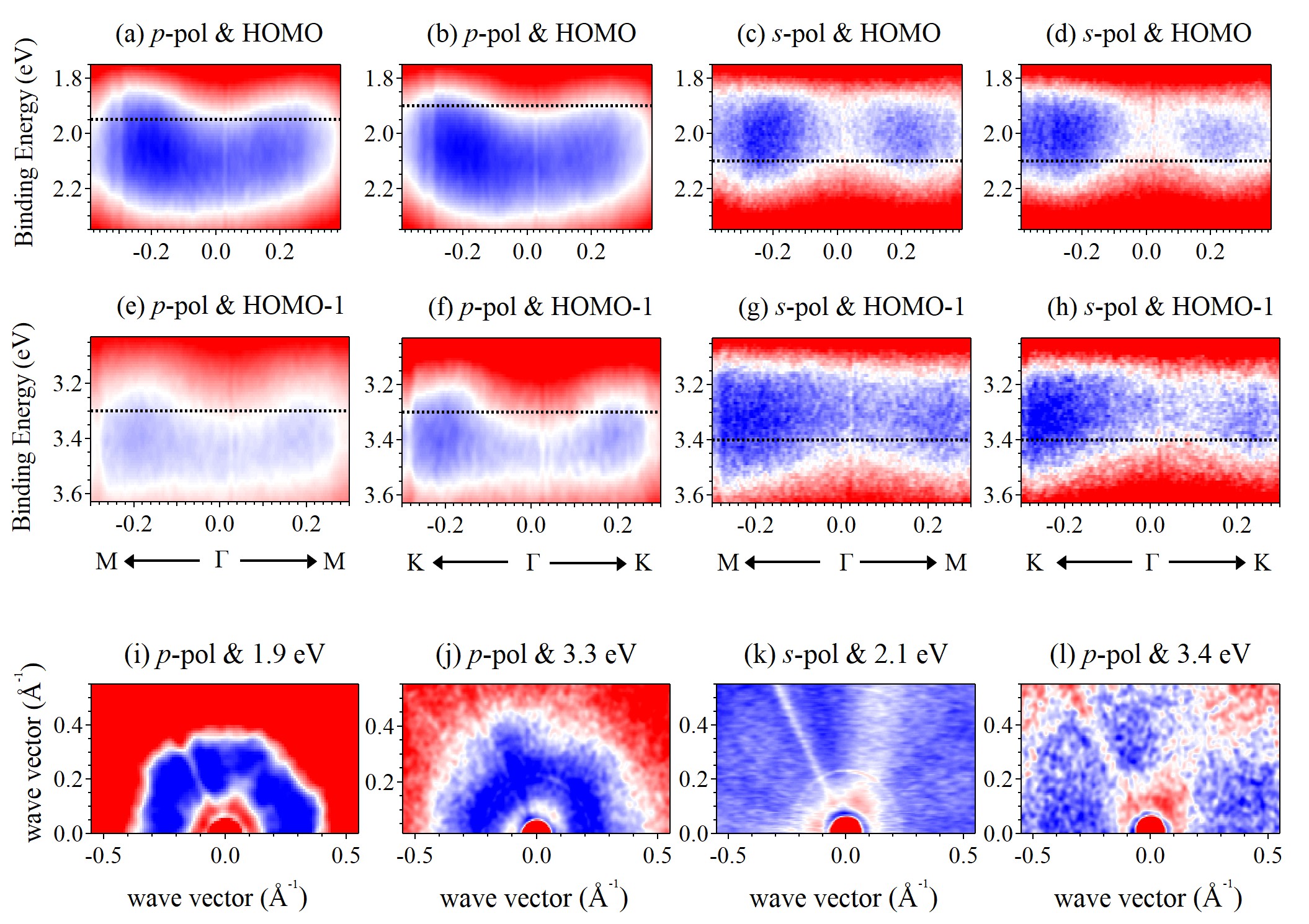}
    \caption{(a-h) ARPES spectra of \csix{}/\bite{} taken around the energy of the HOMO and HOMO-1 bands for $p$-polarized and $s$-polarized light along $\Gamma K$ and $\Gamma M$ directions. (i-l) ARPES equi-energy maps of the HOMO and and HOMO-1 bands for p-polarized and s-polarized light.}
    \label{fig:ARPES2}
\end{figure*}

Let us now move to a detailed study of the molecular energy band dispersion of \csix{} derived bands with HOMO and HOMO-1 character. In the isolated molecule, these orbitals are ten-fold degenerate but in the solid this degeneracy is lifted.~\cite{millie96}  In order to disentangle the sub-band structure, we perform ARPES using linearly $s-$ and $p-$polarized synchrotron radiation. The polarized light couples preferentially to one of the several sub-bands and allows us to probe the sub-bands of HOMO and HOMO-1 levels individually. Figures~\ref{fig:ARPES2}a,b and \ref{fig:ARPES2}c,d show the ARPES spectrum of \csix{}/\bite{} measured in the vicinity of the HOMO and HOMO-1 energy levels employing $p(s)$-polarized light along $\Gamma$M/$\Gamma$K-directions, respectively. Both the HOMO and HOMO-1 bands reveals a significant band-dispersion due to long-range crystalline ordering. A clear electron- and hole-like parabolic dispersion relation can be observed with the $p$- and $s$-polarized lights, respectively. This reveals presence of at least two oppositely dispersive bands at both HOMO and HOMO-1 energy levels. The measured dispersion band width of the HOMO and HOMO-1 bands with $p$-polarized light source are $\sim$ 0.2 and 0.1 eV, respectively. Interestingly, the hole-like band observed with the $s$-polarized light source are weakly dispersive. It is clear that bands having dominant orbital character defined parallel to the interface plane are more dispersive than the out-of-plane direction. This is expected from the band-narrowing effect along the out-of-plane direction due to absence of more than one \csix{} molecules along the out-of-plane direction.

Figure~\ref{fig:ARPES2}i-l shows the 2D ARPES constant energy contours measured in the vicinity of HOMO and HOMO-1 energy levels using linearly polarised photons. In addition to the dispersion of the \csix{} bands, the 2D ARPES constant energy maps reveals an angular anisotropy for both HOMO and HOMO-1 bands. The hexagonal symmetry of constant energy ARPES maps measured using $p$ polarisation confirms the angular anisotropy of both \csix{} HOMO and HOMO-1 dispersive bands. Despite the bands measured using $s$-polarised light are weakly dispersive, the hexagonal anisotropy of the constant energy ARPES contours could be seen in Figure~\ref{fig:ARPES2}k-l. Clearly, the ten-fold degeneracy of the \csix{} is splitted into at least two energy levels due to highly ordered hexagonal crystalline structure on \bite{}. In addition, the equi-energy contours also show hexagonally anisotropic dispersive bands originating from the molecular orbitals.

\subsection{Structural phase transition in \csix{}/\bite{}: electronic and optical properties}
Raman spectroscopy of bulk \csix{} evidences a structural phase transition from fcc to sc around 250~K which is a result of rotational ordering of \csix{}.~\cite{millie96} In the sc structure of bulk \csix{}, all the \csix{} molecules have the same orientation. At low temperatures in the sc bulk \csix{} phase, the Ag(2) Raman mode of \csix{} increases in intensity, shifts to higher frequency and drastically reduces its linewidth.~\cite{millie96} In the following we investigate this phase transition in the limit of 0.5 and one ML \csix{}/\bite{}. Figure 3a depicts the Raman spectra recorded at 300~K and at 80~K for a clean \bite{}, one ML of \csix{}/\bite{} and 0.5 ML of \csix{}/\bite{}. It can be seen that \bite{} has a broad background at 1450~cm$^{-1}$ from which the Ag(2) Raman peak of \csix{} emerges. At a temperature of 85~K, the \csix{} Raman spectra in the vicinity of the Ag(2) mode drastically change. The \csix{} Ag(2) Raman mode shifts to higher frequency, becomes sharper and more intense. The behaviour of the Raman modes in \csix{}/\bite{} upon cooling is very similar to the fcc to sc transition observed in bulk \csix{}.~\cite{Matus1991,millie96} However, as we will show in the following, the details of the structural transition in rotational order of monolayer \csix{} on \bite{} is somewhat different from that observed in bulk \csix{}. Below we show via ARPES that this structural transition in rotational order of \csix{} goes hand in hand with a change in the charge transfer between \csix{} and \bite{}. Figures 3b and 3c depict ARPES of a \csix{}/\bite{} heterostructure recorded at 30~K in the energy range of the Dirac cone of \bite{} and the molecular bands of \csix{}, respectively. Figure 3d shows a comparative MDC analysis of the ARPES spectra of pure \bite{} and \csix{}/\bite{} at RT and \csix{}/\bite{} at 30~K carried out on the same sample. This \bite{} surface was prepared on a different \bite{} film than the surface shown in Figure 1 and therefore has a slightly different carrier concentration. The analysis in Figure 3d affirms the previous conclusion from Figure 1, namely that upon \csix{} deposition, the Fermi surface size of \bite{} shrinks. From Figure 3d we also observe that, upon cooling the \csix{}/\bite{} heterostructure to 30~K, the Fermi surface size increases almost back to the original value of pristine \bite{}. We relate this observation to a temperature-dependent charge transfer between \csix{} and \bite{}. Figure 3e depicts the EDCs of the ARPES data from Figure 3d from \csix{}/\bite{} at RT and at 30~K. The two intense peaks in the energy range shown correspond to the HOMO and the HOMO-1 of \csix{} at two different momenta. It can be seen that the HOMO and HOMO-1 bands shift to higher binding energy by 60~meV and by 110~meV, respectively when cooling down the sample from RT to 30~K. Importantly, this shift is not related to electron doping of \csix{} because calculations below show that the LUMO (not visible in ARPES) also shifts to a higher energy when the sample is cooled. That is, the band gap between HOMO and LUMO for a \csix{} film increases as we cool from 300~K to 30~K. As a consequence, we have charge neutrality of the \bite{}-\csix{} heterostructure despite the electron transfer from \csix{} to \bite{} upon cooling from 300~K to 30~K.

In order to rationalize the observed findings (i.e. the change of charge transfer between \csix{} and \bite{} as a function of temperature), we perform calculations of the total energy and the electron affinity $E_A$ by DFT. Note that $E_A$ is defined as the difference of the total energy between the charged \csix{} and the neutral \csix{}. For the case of strong electron acceptors as \csix{}, the charged \csix{} (with one extra electron in the LUMO) has a lower total energy which leads to negative values of $E_A$. We model the \csix{} ML by considering a \csix{} dimer and calculate its total energy and $E_A$ as a function of the relative angle between the two \csix{} molecules that form the dimer. The dimer is an excellent qualitative approximation to our system in which the \csix-\csix{}  interactions dominate over the \csix-\bite{}  interaction. As a consequence, their electronegativity depends on the relative rotation of neighbouring \csix.
Figure 3f depicts the total energy of the \csix{} dimer as a function of the rotation angle between them.  There is one global minimum at a relative rotational angle of about 70$^\circ$ ($\alpha$ phase). Another local minimum is located at  a relative rotational angle of about 20$^\circ$ ($\beta$ phase). We therefore expect that at low temperatures, all $\csix{}$ are in the $\alpha$ phase and that at increased temperature some of the \csix{} molecules occupy the $\beta$ phase. Let us consider the electron affinity $E_A$ of these two phases in order to rationalize the observed changes in charge transfer to the $\bite$ substrate. Figure 3g depicts the calculated $E_A$ as a function of relative rotational angle of the two \csix{} molecules to one another. For the $\alpha$ phase, the $E_A=-3.123$~eV and for the  $\beta$ phase, $E_A=-3.126$~eV. The smaller value of $E_A$ for the $\beta$ phase means that in this phase, the energy gain per \csix{} is larger, if an electron is accepted from \bite{}. That is, if more \csix{} dimers are in the $\beta$ phase, the electron transfer from \bite{} to \csix{} will be larger. This nicely explains the experimental observation that, at 300~K we have a smaller \bite{} Fermi surface than at 30~K. The calculations were done for a \csix{} dimer and that the calculated difference in $E_A$ between $\alpha$ and $\beta$ phases is 3~meV per \csix{}. Yet, in the experiment, each \csix{} has six neighbours and the total energy difference will be enhanced compared to the calculated value which holds for the \csix{} dimer.

The changes in charge transfer and $E_A$ due to rotational ordering at low temperature result also profoundly affect the optical properties of the heterostructure. To that end, we have recorded the changes of the photoluminescence of \csix{}/\bite{} upon cooling. Figure 3h depicts PL spectra at different temperatures between room temperature and 80~K. It can be seen that upon cooling the luminescence at 1.7~eV increases in intensity. We observe the appearance of PL at temperatures below $\sim$250~K. This is likely related to the increased rotational order which goes together with the absence of electron transfer from the \bite{} to \csix{}. The PL increase at low temperatures may partially be a result of the supressed trion formation in charge neutral \csix{} molecules which decay non-radiatively and due to other supressed non-radiative decay mechanisms. The luminescence spectra of \csix{} are modified by electron-phonon coupling. That is, the luminescence has side peaks that are red-shifted from the main peak by one or several phonon energies (magnified spectrum in Figure 3b). This observation for the case of ML \csix{} is not surprising. Indeed, a strong electron-phonon coupling in bulk \csix{} has been oberved~\cite{Reber1991} and is closely related to superconductivity in alkali metal doped \csix{} such as $Rb_3C_{60}$.
Finally, we summarize the results obtained in Figure 3i showing $E_A$ as a function of temperature. At 30~K, only the $\alpha$ phase contributes and dimers in this phase have a high $E_A$. As a consequence of the high $E_A$, it is energetically not favorable to transfer an electron from \bite{} to \csix{}. At 300~K, also the $\beta$ phase contributes which lowers the effective $E_A$. Thus, it becomes energetically more favorable to transfer an electron from \bite{} to \csix{}. This simple picture neatly explains the experimental observations based on the rotational ordering of \csix{} on \bite{}. It highlights that control of \csix{} rotational order is a means to control charge transfer and optical properties of the \csix{}-\bite{} moir\'e heterostructure.

\begin{figure*}[ht]
    \centering
    \includegraphics[width=15cm]{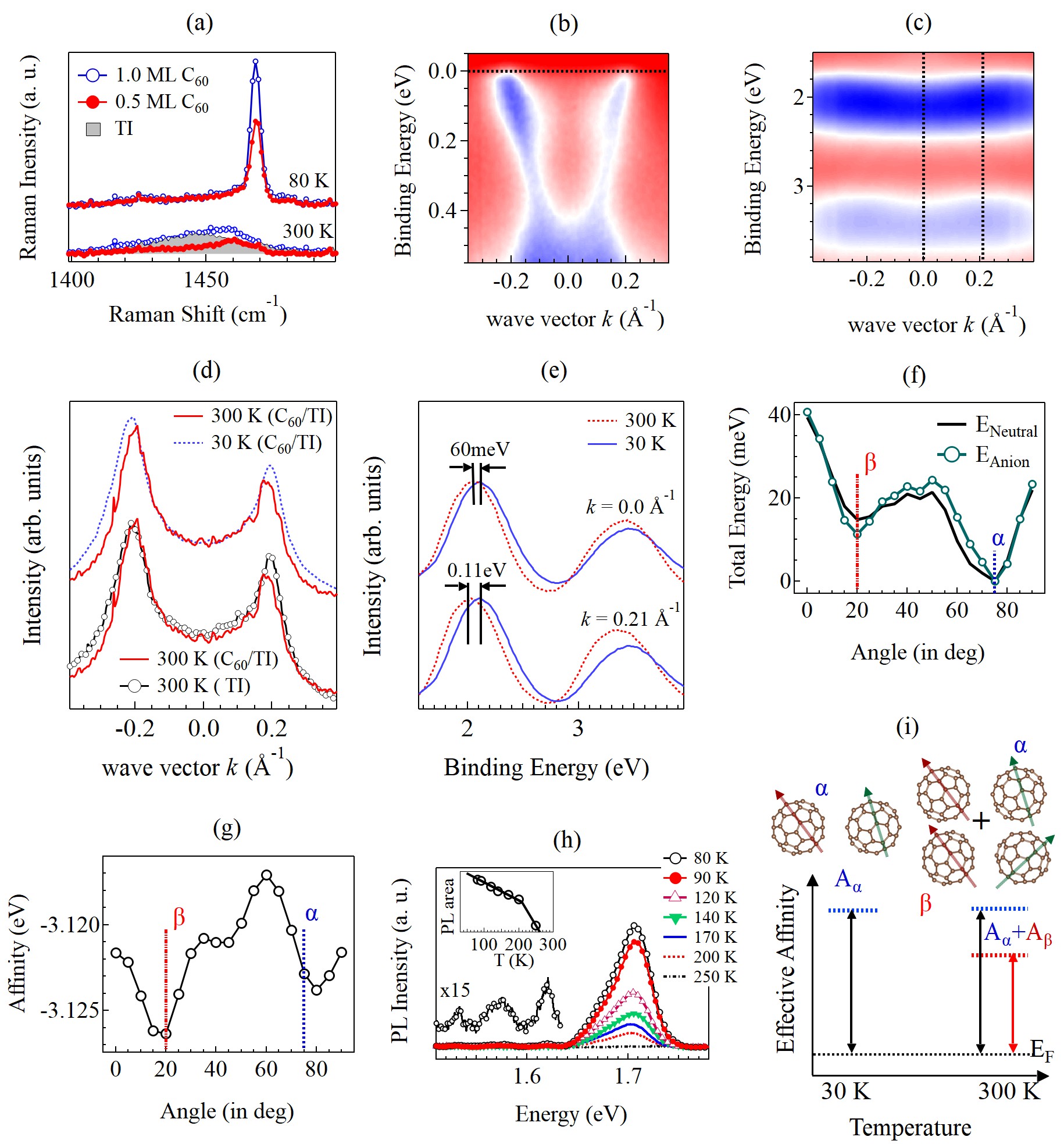}
    \caption{(a) Raman spectra ($\lambda_{exc}=532$~nm) of \csix{}/\bite{} at 80~K and at 300~K for 0.5~ML and 1~ML \csix{} coverages (the \csix{} Ag(2) mode is shown). (b,c) ARPES of \csix{}/\bite{} at 30~K in the energy regions of the TI surface state and the HOMO/HOMO-1 states. (d) MDCs of \bite{} at 300~K and \csix{}/\bite{} at 300~K and 30~K. (e) EDCs of \csix{}/\bite{} through HOMO and HOMO-1 bands at two wavevectors measured at 30~K and 300~K. (f) Total energy of a \csix{} dimer (neutral and anion) vs. the rotation angle between the two constituent \csix{}. Angles $\alpha$ and $\beta$ denote the global and local minima of the total energy. (g) Electron affinity of the \csix{} dimer vs. the relative rotation angle. (h) Temperature dependent photoluminescence (PL) spectra of \csix{}/\bite{} ($\lambda_{exc}=532$~nm). The inset shows the integrated PL intensity. The magnified region shows the phonon side peaks. (i) Schematic diagram of dimers with angles $\alpha$ and $\beta$ between \csix{}. At low temperature, all dimers are in the ground state with relative rotational angle $\alpha$ between \csix{}. At high temperatures, the state with angle $\beta$ between the \csix{} becomes populated. At angle $\beta$, the value of the affinity is lower explaining why \csix{} are ionized more easily at higher temperature. }
    \label{fig:Raman}
\end{figure*}

\section{Conclusion and outlook}
We have synthesized a weakly interacting interface between \bite{} and a highly ordered monolayer \csix{} 
and investigated its structural, electronic and optical properties. We have found via LEED measurements that \csix{}
crystallizes with long-range order that is different from thick \csix{} films on TI and has not yet been reported for a TI--organic interface. We have observed a structural phase transition in the rotational order of monolayer $\csix$ that is reminiscent to the structural transition in bulk \csix{}. However, in the case of ML \csix/\bite{}, theory suggests a different rotational angle for neighbouring \csix{} in the low temperature phase than for bulk \csix{}. This rotational ordering goes hand-in-hand with a modified charge transfer. Our work illustrates the complex temperature-dependent interplay between charge transfer and rotational order in heterostructures between \csix{} and TI. Our observations are explained by DFT calculations of the total energy and electron affinity  of $\csix{}$. Our work highlights that topological insulator / organic heterostructures are interesting material systems with tunable optoelectronic properties that can be grown by van-der-Waals epitaxy. We expect that similar results may be achieved for 2D material / organic heterostructures. This research direction could e.g. be followed up in one of the following ways. First, magnetic fullerenes on TI can be grown as a material system to study the effects of Zeeman splitting on the spin-polarized bands of the TI. Second, the van der Waals epitaxy of $\csix{}$ might be extended to superconducting alkali metal doped fullerene films to form novel superconductor - TI interfaces that can be grown by vdW epitaxy. Third, this material system might be used as a substrate for the growth of another highly ordered organic film anchored to the \csix{} molecules.

\section{Methods}
\bite{} films of 20\,nm thickness were grown on a n-doped Si(111) wafer using MBE
in a UHV chamber with base pressure <\,\num{5e-10}\,mbar. Before exposure to air, the films were capped with an 
amorphous 2\,nm Al$_2$O$_3$ films. In a separate preparation chamber of the ARPES system with a base pressure <\,\num{5e-10}\,mbar,
the capping layer was removed by means of sputtering and annealing cycles with temperatures as high as 650\,K until no improvement of the LEED spectrum was distinguishable any more. \csix{} ($>\,95.5\%$ HPLC) was evaporated
from a Knudsen cell at 675\,K with an evaporation rate of 0.5\,\AA{}min$^{-1}$ as calibrated
by a quartz microbalance (QMB). During evaporation, the substrate temperature was held at 400\,K
and post-annealed at 420\,K for 30\,min. ARPES measurements in Figure 1 were performed at the laboratory using an MBS A1 analyzer and a He gas discharge lamp. Data shown in Figures 2 and 3 were recorded on an identically prepared sample with a slightly different carrier concentration. ARPES spectra in Figures 2 and 3 were measured using synchrotron radiation with a photon energy of 21\,eV at the BaDElPh beamline (ELETTRA).~\cite{Petaccia2009} Raman and PL spectra shown in Figure 3 were measured on the identical sample as ARPES data from Figures 2 and 3 inside an optical cryostat. During Raman and PL measurements, the vacuum inside an optical cryostat was better than $10^{-6}$~mbar and a green laser (532~nm) with a laser power of 5~mW was used.

\begin{acknowledgement}
  A.G. and N.A. acknowledge the Deutsche Forschungsgemeinschaft (DFG) through CRC 1238 (project no. 277146847, subprojects A01 and C01). B.S. and A.G. acknowledge project SE 2575. A.G. acknowledges DFG for the project INST 216/955-1 FUGG. The stay at the Elettra synchrotron for ARPES experiments has been supported by the project CALIPSOplus under Grant Agreement 730872 from the EU Framework Programme for Research and Innovation HORIZON 2020. We acknowledge Elettra Sincrotrone Trieste for providing access to its synchrotron radiation facilities. The authors gratefully acknowledge the computing time granted by JARA Vergabegremium and provided on the JARA Partition part of the supercomputer JURECA at Forschungszentrum Jülich.
  
\end{acknowledgement}

\bibliography{c60TIbib}
\end{document}